\documentclass[superscriptaddress,amsmath,amssymb,aps,prd,twocolumn]{revtex4-2}

\usepackage{color}
\usepackage{amssymb}
\usepackage{epsfig}
\usepackage{graphicx,color}
\usepackage[usenames,dvipsnames]{xcolor}
\usepackage[section]{placeins}
\usepackage{amsmath}
\usepackage{subfigure}
\usepackage[normalem]{ulem}
\usepackage{booktabs}
\usepackage{xcolor}
\usepackage{epsfig,graphics,color,graphicx,amsmath}

\colorlet{darkgreen}{green!50!black}
\colorlet{brightyellow}{yellow!75!red}
\colorlet{orange}{red!50!yellow}
\colorlet{darkblue}{blue!60!black}
\colorlet{darkred}{red!80!black}

\def\bwt{\begin{widetext}}
\def\ewt{\end{widetext}}

\newcommand{\be}{\begin{eqnarray} &&}
\newcommand{\ee}{\end{eqnarray}}
\newcommand{\nonu}{\nonumber \\ & &}

\def\psla{\slash \! \! \!}

\usepackage{soul}

\usepackage{mathtools}
\usepackage{mathrsfs}
\usepackage{bm}
\usepackage{relsize}
\usepackage{indentfirst}

\def\slashed{\slash \! \! \!}

\usepackage{xcolor}

\begin{document}
\title{The chiral limit of a fermion-scalar $(1/2)^+$ system in covariant gauges}

\author{A.~Noronha}
\affiliation{Instituto Tecnol\'ogico de Aeron\'autica,  DCTA,
12228-900 S\~ao Jos\'e dos Campos,~Brazil}
\author{W. de Paula}
\affiliation{Instituto Tecnol\'ogico de Aeron\'autica,  DCTA,
12228-900 S\~ao Jos\'e dos Campos,~Brazil}
\author{J.H. de Alvarenga Nogueira}
\affiliation{Instituto Tecnol\'ogico de Aeron\'autica,  DCTA,
12228-900 S\~ao Jos\'e dos Campos,~Brazil}
\author{T.~Frederico}
\affiliation{Instituto Tecnol\'ogico de Aeron\'autica,  DCTA,
12228-900 S\~ao Jos\'e dos Campos,~Brazil}
\author{E. Pace}
\affiliation{Universit\`a di Roma ``Tor Vergata'', Via della Ricerca Scientifica 1, 00133 Rome, Italy}
\author{G. Salm\`e}
\affiliation{
INFN, Sezione di Roma, P.le A. Moro 2, 00185 Rome, Italy}

\date{\today}

\begin{abstract}
The homogeneous Bethe-Salpeter equation (BSE) of  a (1/2)$^+$ bound system, that has both fermionic and bosonic degrees of freedom, that we call a {\em mock nucleon}, is studied in Minkowski space, in order to analyse the chiral limit in  covariant gauges. After adopting an interaction kernel  built with a one-particle exchange,  the $\chi$-BSE  is numerically solved by means of the Nakanishi integral representation and light-front projection. Noteworthy,  the chiral limit induces a scale-invariance of the model and consequently generates a wealth of striking features:  i) it reduces the number of non trivial Nakanishi weight functions  to only one; ii) the form of the surviving  weight function has a factorized dependence on the two relevant variables, compact and non-compact one; iii)  the coupling constant becomes an explicit   function of the real exponent governing  the power-law fall-off of the non trivial Nakanishi weight function.   The thorough investigation at  large transverse-momentum of light-front Bethe-Salpeter amplitudes,  obtained  with massive constituents,  provides a  confirmation of  the  expected universal power-law fall-off, with   exponents predicted by our non-perturbative framework. Finally, one can shed light on the exponents {that} govern the approach to the upper extremum of the longitudinal-momentum fraction  distribution function of the {\em mock nucleon}, when the coupling constant varies.

\end{abstract}
\maketitle

\section{Introduction}\label{Sec:intr}
Quantum chromodynamics (QCD) in Euclidean lattice has achieved a massive success in describing hadronic properties,  but nonetheless accurately investigating dynamical quantities, e.g. by a direct calculations of the parton distributions (see, e.g., Refs. \cite{Constantinou:2020hdm,HadStruc:2021qdf,Ji:2022ezo} and references therein) or the gluon role in the emergent hadronic masses (see, e.g. Ref. \cite{Roberts:2021nhw} and references therein), is still a challenge.  Therefore, developing phenomenological but rigorous tools in Minkowski space for investigating hadronic dynamical properties could help to offer  complementary insights, drawn directly from the physical space, and help to accumulate a critical mass of information on the non-perturbative regime of the strong dynamics. Plainly,  such insights could  be relevant for the analysis of the  unprecedented amount of  accurate data on the hadron structure that will be gathered at the future Electron Ion Colliders~\cite{AbdulKhalek:2021gbh,Anderle:2021wcy}.

A phenomenological way to treat the nucleon is to consider a constituent quark point of view, i.e.  the dressing of the light degrees of freedom (dof) through an effective mass  generated by the gluon non-perturbative interaction, so that    the chiral symmetry is substantially broken. For instance, following Refs. \cite{Alkofer:2000wg,Eichmann:2016yit}, one can see  how a quark-diquark description of the nucleon   emerges  by  introducing a simple two-step process applied to the relativistic three-body bound-state equation (Faddeev-Bethe-Salpeter equations), namely by neglecting the three-body forces and approximating  the quark–quark scattering matrix as a sum over separable diquark correlations. In particular, the interaction kernel of the coupled equations, that determine  the Faddeev components of the Bethe-Salpeter (BS) amplitude, highlights the quark-exchange mechanism, without an explicit presence of gluons.  Therefore, the  emergent picture of the nucleon as a quark-diquark should be viewed as a useful  tool  to  deepen the understanding   of the effective dynamics driving the relative motion in each Faddeev amplitude,  with  a given spectator  quark.
Indeed, the diquark correlation has an important role in the hadron structure, particularly in view of the recent discoveries of multi-quark states (see e.g., Ref. \cite{Barabanov:2020jvn} for a review). It is also relevant to mention that lattice QCD (LQCD) simulations showed that diquark correlations emerge from the QCD dynamics~\cite{Alexandrou:2006cq,DeGrand:2007vu,Babich:2007ah, Bi:2015ifa,Francis:2021vrr,Watanabe:2021oyv,Watanabe:2021nwe}.

Recently a $(1/2)^{+}$ bound-state system, composed by a pair of massive fermion  and  scalar boson interacting through a vector-boson exchange, 
namely a {\em mock nucleon},  was  investigated~\cite{AlvarengaNogueira:2019zcs} by using the homogeneous  Bethe-Salpeter equation (BSE) in Minkowski space and the Nakanishi integral representation (NIR) of the BS amplitude (see Refs. \cite{Nakanishi:1963zz,Nakanishi:1971}).  In this dynamical model,  the fields represent a quark and a  structureless scalar-diquark, interacting through an effective massive one-gluon-exchange,  that implements the   non-perturbative dressing of the gauge-field quanta as it has been obtained by  LQCD calculations in the Landau gauge~\cite{Dudal:2013yva}. 
Interestingly, the model has a dimensionless coupling constant and, therefore, one expects that it would be scale invariant in the chiral limit. This remarkable  feature has far-reaching  consequences for the BS amplitude, which will be explored in our work by also considering the dependence on the  covariant gauges. In particular, through the solutions of the BSE we aim to gain insight into   the stability of the fermion-boson system, that imposes a gauge-dependent  upper-bound for the coupling constant, in the spirit of the Miransky scaling~\cite{Miransky:1984ef,Miransky:1996pd,Miransky:2000rb}, that enlights the role of the conformal symmetry in the description of the non-perturbative regime of the gauge theories.
Also in other systems, like  fermion-fermion or fermion-antifermion bound states, the transition from stable solutions of the BSE to unstable ones as the coupling constant approaches a critical value,  occurs
both in Euclidean~\cite{Dorkin:2007wb} and in Minkowski spaces~\cite{Mangin-Brinet:2001rtp,Mangin-Brinet:2001wmj,Carbonell:2010zw}.  It should be pointed out that such  a sharp transition, triggered by the loss of  scale invariance  can be found in several areas (see, e.g., Refs.~\cite{Miransky:1996pd,Kaplan:2009kr,Enss:2010qh,Nakayama:2013is}). The instability in the  solution of the BSE is analogous to the {\em Landau fall to the center}~\cite{LandauQM} and represents the breaking of the continuous scale symmetry to a discrete one. This phenomenon is present in the Thomas collapse of the non-relativistic three-boson system in the limit of zero-range interactions~\cite{Thomas:1935zz} and in the Efimov effect~\cite{EFIMOV1970563,EFIMOV198145}, unified as the Thomas-Efimov effect (see e.g.~\cite{Frederico:2012xh}). 
 
Our aim is to explore the  solutions, in the chiral limit, of the fermion-scalar ladder BSE, obtained by using the NIR of the BS amplitude and the so-called LF projection, that amounts to integrate over the minus component of the LF-momentum $k^-=k^0-k^3$ (see the application of this technique  to systems with only fermionic dof  in Refs.~\cite{dePaula:2016oct,dePaula:2017ikc}). The analysis in the chiral limit of the system of integral equations for the Nakanishi weight functions (NWFs), formally derived from the initial homogeneous BSE once the NIR is adopted, allows one to highlight nontrivial outcomes, like: i)  the existence of a critical value of the coupling constant; ii) the explicit relation between coupling constant and  the power governing the universal power-like fall-off of the transverse-momentum distribution; iii) the prediction of the exponent controlling the longitudinal-momentum fraction distribution, at the largest end-point; iv) the comparison with calculations of 
transverse-momentum distribution for massive constituents in the ultra-violet (UV) region, namely where the  constituent masses can be disregarded. It has to be emphasized that  also the dependence of the coupling constant upon covariant gauges has been   quantitatively investigated.

It should be recalled that the main advantage of disposing of such a dynamical model, albeit a  simple one, is that a study can be carried out  in a formally exact framework.  In this way, the assumptions are clearly stated and  the properties of the solutions, as well as  of  the associated distributions,  can be traced back  to the features of the physics content of the interacting kernel and the symmetries of the  dynamical equations governing the model.

This work is organized as follows. In Sect.~\ref{sect:BSE}, it is presented  the fermion-scalar homogeneous  BSE model in covariant gauges and the NIR of the scalar amplitudes that allows to expand the BS amplitude on the relevant operators for $1/2^+$ system. In Sect.~\ref{sect:chi}, the uncoupled integral equations for the Nakanishi weights functions are obtained by exploring the 
scale invariance of the homogeneous BSE in the chiral limit. In Sect.~\ref{sect:numerical}, the results obtained from the
numerical solutions are provided and the results discussed. In Sect.~\ref{Sect:summary}, our conclusions are drawn.

\section{BSE and the Nakanishi integral representation}\label{sect:BSE}
The bound  system under consideration  is governed by the following interacting Lagrangian~\cite{AlvarengaNogueira:2019zcs}
\begin{equation}
\begin{aligned}\label{LagFerBos}
{\mathcal L}=& \lambda_F \, \bar \psi \, \slashed{V} \, \psi  - i\lambda_S \, \phi^*\overleftrightarrow{\partial}_\mu\phi \, V^\mu,
\end{aligned}
\end{equation}
where $V^\mu$ is the interacting vector-boson field, $\phi$ and $\psi$ are the scalar and fermionic fields, respectively, and the coupling constants are dimensionless. In~\cite{AlvarengaNogueira:2019zcs}, this system was studied in the Feynman gauge. Here we extend the analysis for an arbitrary covariant gauge, in the chiral limit.
The homogeneous BSE for a fermion-boson system, forming a  $(1/2)^{+}$ bound state, with ladder vector interaction is
\begin{small}
\begin{equation}
\Phi(k,p) = G_{0} (k_s) ~ S(k_f)\int \frac{d^4 k'}{(2\pi)^4} \,iK(k, k', p)~\Phi(k',p) \, ,\label{eqBSBF}
\end{equation}
\end{small}
where $\Phi(k,p) $ is the BS amplitude, $G_0(k_s)$ is the scalar propagator, $S(k_f)$ is the fermionic propagator
\begin{equation}
G_{0} (k_s) = \frac{i}{k_s^2 -m_{s}^2 + i\epsilon} ~ , \quad
 S(k_f) = i\frac{ \psla k_f + m_{f}} { k_f^2-m_{f}^2+i\epsilon} \label{propSF} \, ,
\end{equation}
with the scalar and fermion momenta given by $k_s = p/2 - k$ and $k_f = p/2 + k$, respectively. In addition, $i K(k, k', p)$ is the interaction kernel, that reads in a particular covariant gauge $\zeta$
\begin{align}
&i K(k, k', p) =  - i \, \lambda_S \, \lambda_F~\gamma^\mu~\frac{ (p -  k -  k')^\nu}{(k - k')^2 - \mu^2 + i\epsilon} \nonumber\\
& \times \left[g_{\mu\nu}-(1-\zeta)~\frac{(k^{\prime} - k)_\mu \, (k^{\prime} - k)_\nu}{(k-k')^2-\zeta \, \mu^2+i\epsilon}\right] \, ,
\label{propD}
\end{align}
where $\mu$ is the vector-boson mass.

The BS amplitude for a $(1/2)^{+}$ system can be decomposed in the following Dirac basis \cite{AlvarengaNogueira:2019zcs}
\begin{equation}
\Phi(k,p)=[O_1(k)\phi_1(k,p)+O_2(k)\phi_2(k,p)]U(p,s) \, ,
\label{diracbasis} 
\end{equation}
where $\phi_i (k,p)$ are scalar functions, $U(p,s)$ is the spinor of the bound state with squared mass $M^2 = p^2$, normalized as $\bar{U} U = 1$, satisfying $\left(\psla p - M \right) U(p,s) = 0$, with $O_1(k) = \hat{1}$ and $O_2(k) = \psla k /M$ being operators. The mass of the system $M$ is related to the binding energy $B$ as $M= 2\bar{m} - B$, where $\bar{m} = (m_f + m_s)/2$.

An important tool to solve the BSE in Minkowski space is the NIR. It provides the analytical structure in terms of the external momenta. With this in mind, we express each component of the BS amplitude as follows~\cite{Nakanishi:1963zz,Nakanishi:1971}
\begin{small}
\begin{equation} 
\phi_i(k,p) = \int_{-1}^{+1}\hspace{-.1cm}dz'\int_{0}^{\infty}\hspace{-.1cm}d\gamma'\frac{g_i(\gamma',z')}{\left[k^2 + p\cdot k \, z'- \kappa^2 -\gamma'+i\epsilon\right]^3} \label{RINka}\, ,
\end{equation}
\end{small}
where $g_i(\gamma',z')$ are the NWFs and $\kappa^2 = \bar{m}^2 - M^2/ 4$. Then, one can  analytically perform both relevant  four-dimensional integrations as well as the critical  LF projection~\cite{dePaula:2022pcb,Ydrefors:2021dwa}. 

Remarkably, by combining the NIR and a LF projection, it is possible to formally obtain a system of coupled integral equation for the NWFs from the BSE, as given for the massive case in  Ref.~\cite{AlvarengaNogueira:2019zcs}. 

\section{Chiral limit  and  scale invariance}
\label{sect:chi}
In order to carry out the chiral limit, one simply puts both the constituent masses, $m_s$  and $m_f$, the vector-boson mass, $\mu$, and the bound-system mass  $M$ equal to $0$. In this limit, the system is decoupled and one remains with the following integral equations:
\begin{align}
\int_{0}^{\infty}&d\gamma'\frac{g_1(\gamma',z)}{[\gamma + \gamma']^2} = - \frac{\alpha}{ 2 \, \pi } \int_{-1}^{+1} dz'\int_{0}^{\infty} d\gamma'\,g_1(\gamma',z') \nonumber\\
& \times 
\left( \mathcal{P}_{1}^{(1)}(\gamma, z, \gamma', z') + (1-\zeta)  \, \mathcal{P}_{1}^{(2)}(\gamma, z, \gamma', z') \right)  \, , 
\label{uvg1}
\end{align}
\begin{align}
 \int_{0}^{\infty}&d\gamma'\frac{g_2(\gamma',z)}{[\gamma + \gamma']^2} = - \frac{\alpha}{ 2 \, \pi } \int_{-1}^{+1}dz'\int_{0}^{\infty} d\gamma'\, g_2(\gamma',z')  \nonumber\\
&\times \left( \mathcal{P}_{2}^{(1)}(\gamma, z, \gamma', z') +(1-\zeta)  \mathcal{P}_{2}^{(2)}(\gamma, z, \gamma', z') \right) \, , 
\label{uvg2}
\end{align}
where $z = - 2 k^{+}/M$, with  $k^+=k^0+k^3$ the longitudinal-momentum, and 
$\alpha = (\lambda_S \, \lambda_F )/(8 \, \pi) $. In particular, the longitudinal-momentum fraction for the fermion is $\xi=(1-z)/  2$ and for the boson is $1-\xi=(1+z)/2$. The functions defining the kernel  $\mathcal{P}_{i}^{(a)}$ are given in Appendix \ref{appendixa}.

The coupling constants, $\lambda_S$ and $\lambda_F$, for the fermion-boson system are dimensionless, as well as $\alpha$. Theories with this property become scale invariant in the chiral limit. In fact, one can explicitly verify the scale invariance of Eqs.~\eqref{uvg1} and \eqref{uvg2} by applying  the scale transformation $\gamma \to \Lambda \, \gamma$, with $\Lambda$ any constant, and recalling that $\gamma'$ is an integration variable, and one can use $\gamma' \to \Lambda \,  \gamma'$.  Therefore one expects that the solutions of Eqs.~\eqref{uvg1} and \eqref{uvg2}   are homogeneous functions in $\gamma$ with a power-law behavior~\cite{CLOS,DePaula:2019ryz, takayasu1990fractals}. This property naturally leads to   the following factorization Ansatz:
\begin{equation}
g_i(\gamma,z)=\gamma^{r} f_{i, r}\, (z) \, .
\label{eq:g2}
\end{equation}
\begin{figure*}[htb]
\begin{center}
\epsfig{figure=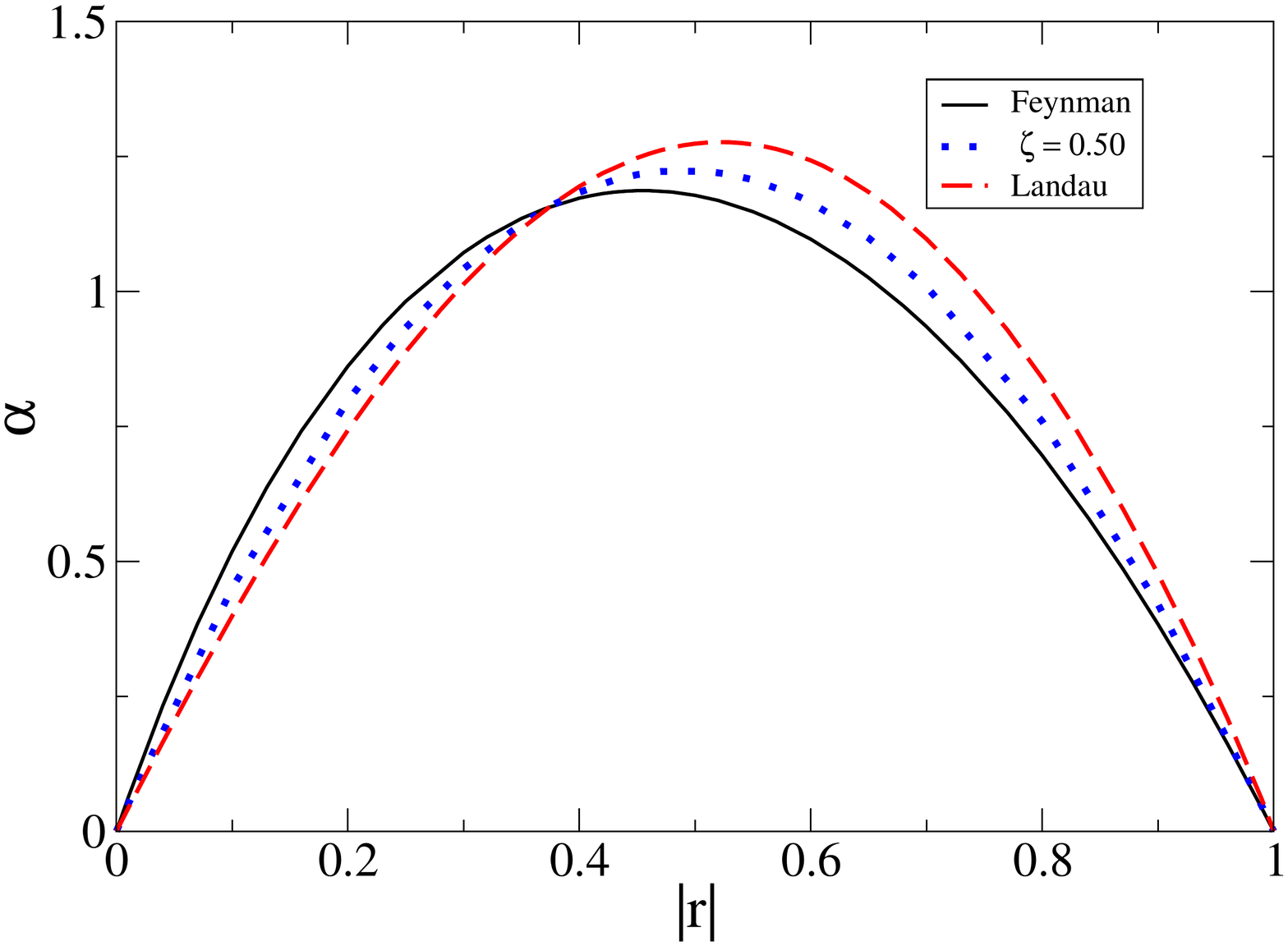,width=9.0cm}
\hspace{-0.5cm}\epsfig{figure=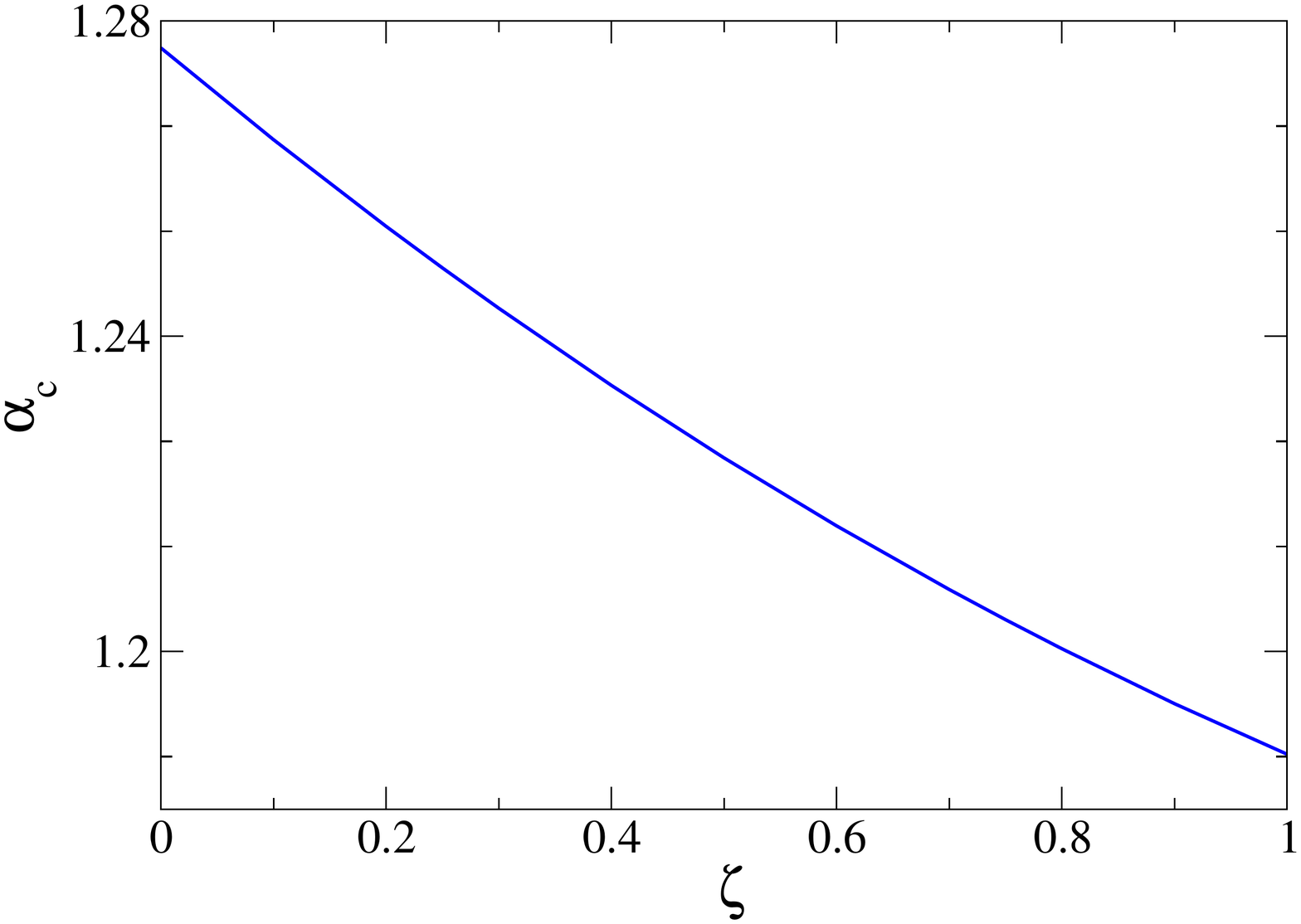,width=9.0cm}
\caption{(Color online). Left panel: coupling constant of the fermion-boson system in the chiral limit, Eq. \eqref{eq:alpha}, vs. the power $r$ (see  Eq. \eqref{eq:g2}). Solid line: Feynman gauge ($\zeta=1$). Dotted line:  $\zeta = 0.5$ gauge. Dashed line: Landau  gauge ($\zeta=0$). Right panel: the critical value  $\alpha_c$ of the coupling constant in Eq. \eqref{eq:alpha} for different gauges in the range $0\le \zeta\le 1$.}\label{Fig1}
 \end{center}
\end{figure*}
After substituting our Ansatz~\eqref{eq:g2} in Eq.~\eqref{uvg1}, one can show that the condition for a finite integral is $0<r<1$. This solution is not physically acceptable because the LF projection of the scalar function $\phi_1$ in the impact-parameter space would have a divergence at the origin (see, e.g., Refs.~\cite{Gutierrez:2016ixt,dePaula:2020qna} for a discussion of the BS amplitude in the impact-parameter space for both  two-scalar bound system and pion, respectively). On the other hand, for Eq.~\eqref{uvg2},  one realizes  that the power $r$ has to fulfill the constraint $-1<r<0$ for getting  a well-defined integration on $\gamma'$ and   avoiding the singularity of the kernel at the end-points (cf.  Eq.~\eqref{eq:BSfz2} below).

Therefore, from now on, one can retain only Eq. \eqref{uvg2}. By performing the integration on $\gamma$  of both sides of Eq. \eqref{uvg2}, one obtains an integral equation for $f_{2,r}(z)$,  and  looks for real eigenvalues $\lambda_2=2\pi/\alpha$. Namely, one remains with the following integral equation:
\begin{equation}
 f_{r}(z) = \frac{\alpha}{2 \, \pi} \hspace{-.05cm} \int_{-1}^{+1} \hspace{-0.2cm} dz' \left[ {G}(z,z') + (1-\zeta)  {H}(z,z')\right]   f_{r}(z') \, ,
\label{eq:BSfz2}
\end{equation}
where the subscript 2 in $f_{r}(z)$ has been dropped out for simplicity and the two contributions to the kernel are given by
\bwt
\be
{G}(z,z') =   \frac{1}{ 2\, |r| (1-|r|)}  \left[ \left( \frac{1+z}{1+z'} \right)^{|r|} \,  \theta(z' - z)   
+  \left(1 + \frac{4 \, |r|}{(1-z')} \right) \left( \frac{1-z}{1-z'} \right)^{|r|} \, \, \theta(z - z') \right] \, ,\label{eq:G1Kernel}
\ee
and
\be
H(z,z') =  \frac{1}{2\, \, |r| \, (1-|r|)~(2-|r|) }  \, \Bigg\{
\left( \frac{1+z}{ 1+z'}\right)^{|r|} ~\theta(z'- z)~
 \Biggl[\frac{-|r|^3 + 7 \, |r|^2 - 11 \, |r| + 3}{ (3-|r|)} + \frac{|r| \, (1-|r|)}{ (1+z')} \Biggr] 
 \nonu
+    \left( \frac{1-z }{1-z' }\right)^{|r|} 
~\theta(z - z') \Bigg[ \frac{-|r|^3 + 7 \, |r|^2 - 11 \, |r| + 3}{(3-|r|)}   
 -  \frac{|r| (1-2\, |r|)}{ (1-z')}
\Bigg] \Bigg\} \, .\label{eq:G2Kernel}
\ee
\ewt
One can explicitly check that the combination of the factor $(1\pm z)$  with the corresponding theta-function yields a vanishing kernel at the end-points ($z = \pm 1$), implying $f_r(\pm 1) = 0$.
 For a finite $r$, the kernel is not symmetric under the transformation $z \leftrightarrow -z$ (this feature entails real and complex conjugated eigenvalues) and therefore we expect that the solutions are non symmetric, i.e. $f_{r}(z) \neq f_{r}(-z)$. Due to this property, it is interesting to observe that in the chiral limit  the momentum fraction distributions of the boson and the fermion are distinct, as illustrated in what follows.

By integrating on $z$ both sides of  Eq. \eqref{eq:BSfz2}, after  inserting the change of variables $y_\pm=(1\pm  z)/(1\pm z')$, one can obtain the following expression of the coupling constant $\alpha$ (see details in Appendix \ref{app_alpha})
\bwt
\be
\alpha (r,\zeta) =   2 \pi \,  \Biggl[ {1+ 2 |r| \over |r| (1-r^2)} 
+ (1-\zeta)  \frac{ - 3 \, |r|^3 + 17 \, r^2 - 22 \, |r| + 6  }{2\, \, |r| \, (1-r^2)~(2-|r|) (3-|r|) }   \Biggr]^{-1}. 
\label{eq:alpha}
\ee
\ewt
The coupling constant $\alpha(r,\zeta)$  
as a function of the power $r$, for different gauges, is presented in the left panel of Fig.~\ref{Fig1}.  At fixed  gauge,  $\alpha(r,\zeta)$ grows for increasing $|r|$ up to a maximum value, that we call {\em critical coupling constant}, $\alpha_c$.  The corresponding critical value of $|r|$ is reached  around $|r_c|\sim 0.5\pm 0.1$ (the uncertainties is given by the variation of $\zeta\in[0,1]$ ).  For $\alpha>\alpha_c$ the power $r$ becomes complex and Eq.~(\ref{eq:BSfz2}) presents a pair of log-periodic solutions, (see, e.g., Ref.~\cite{LandauQM}), which demands one extra scale to determine the solution uniquely: a phenomenon known as Miransky scaling~\cite{Miransky:1984ef,Miransky:1996pd,Miransky:2000rb} in the context of  quantum field-theory.
The critical value $\alpha_c$ depends on the covariant gauge, as shown in the right panel of Fig.~\ref{Fig1}, where the gauge choice has been restricted to the interval $[0,1]$. The value of $\alpha_c$ smoothly decreases  with respect to  $\zeta$ from the Landau to the Feynman gauge, changing  by only $\sim 8\%$. 
The second term in Eq.~\eqref{eq:alpha} leads to a decreasing $\alpha_c$ for $\zeta\to 1$. The analysis of a more general case with $\zeta \not\in[0,1]$, as well as non covariant gauges, is left for a future work. 

It is worth noticing that from the left panel of Fig.~\ref{Fig1} one is able not only to separate  the region where the solutions Eq.~\eqref{eq:BSfz2} are stable, but also where  they are  unphysical. This can be understood by recalling that for {\em an increasing value} of $\alpha$ the binding grows, and hence the system becomes more compact, i.e. the average size of the system  shrinks. The latter feature translates (see Eqs. \eqref{lfwf} and \eqref{lfwfUV} below, for a hint) in a tail of the momentum distribution higher, from a heuristic application of the uncertainty principle. This  eventually means {\em a smaller value} of the power $r$ in Eq.~\eqref{eq:g2}. For the branch where $r< r_c$, one notices from the left panel of Fig.~\ref{Fig1} that the derivative $d\alpha/dr$ is positive, and therefore this region has to be excluded in the analysis of physical systems.

\section{Numerical solutions}
\label{sect:numerical}
The solutions of the integral equation for $f_{r}(z)$ have been  obtained by expanding $f_{r}(z)$ onto a spline basis. In this way one discretizes  the eigenvalue problem, where the coupling constant $\alpha$ is the inverse of the eigenvalue. We solve Eq.~\eqref{eq:BSfz2} looking for the largest real eigenvalue within a set of   $r$ values, in different covariant gauges.

\begin{figure*}[thb]
\begin{center}
\epsfig{figure=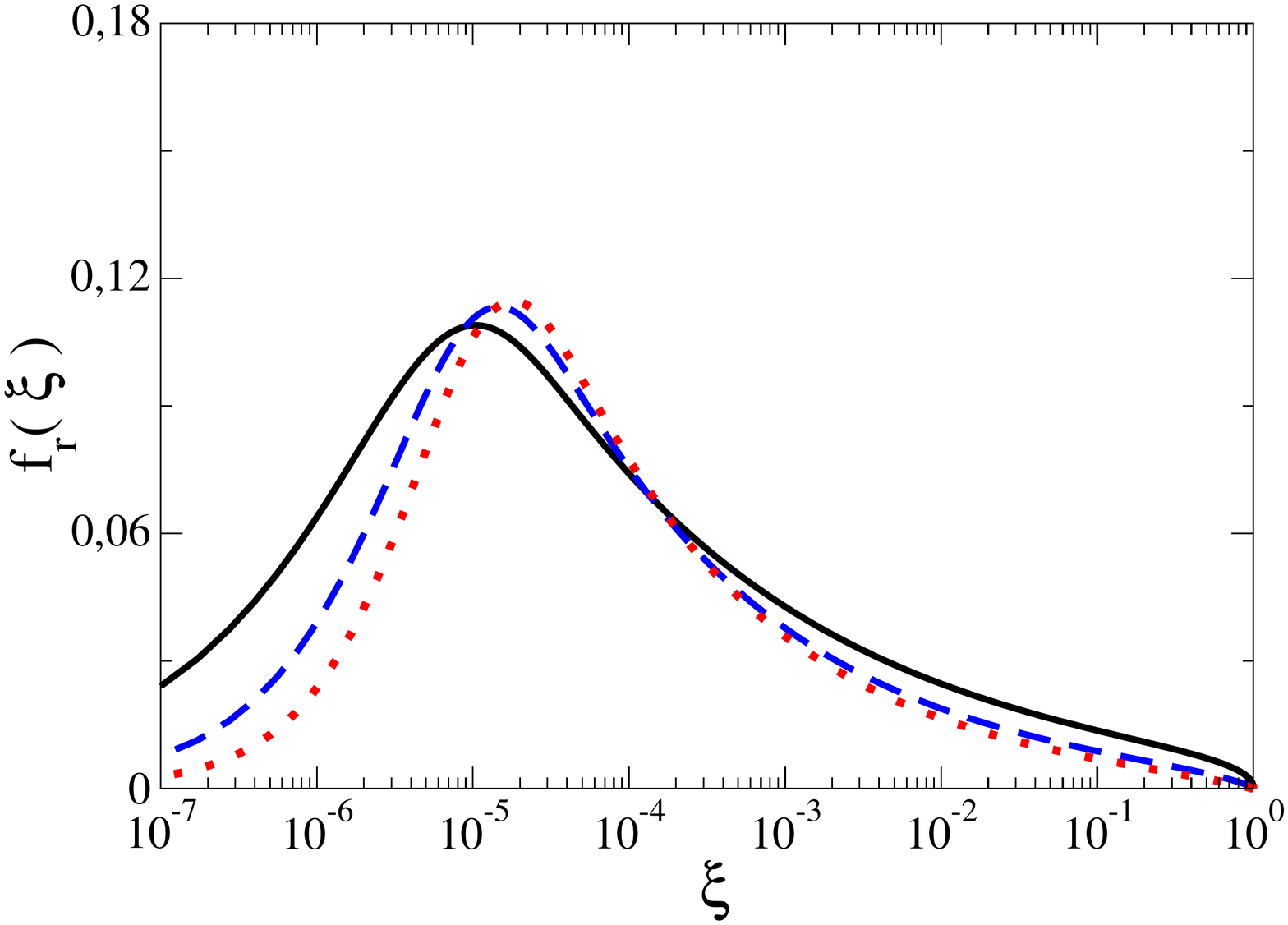,width=9.0cm}
\hspace{-0.5cm}\epsfig{figure=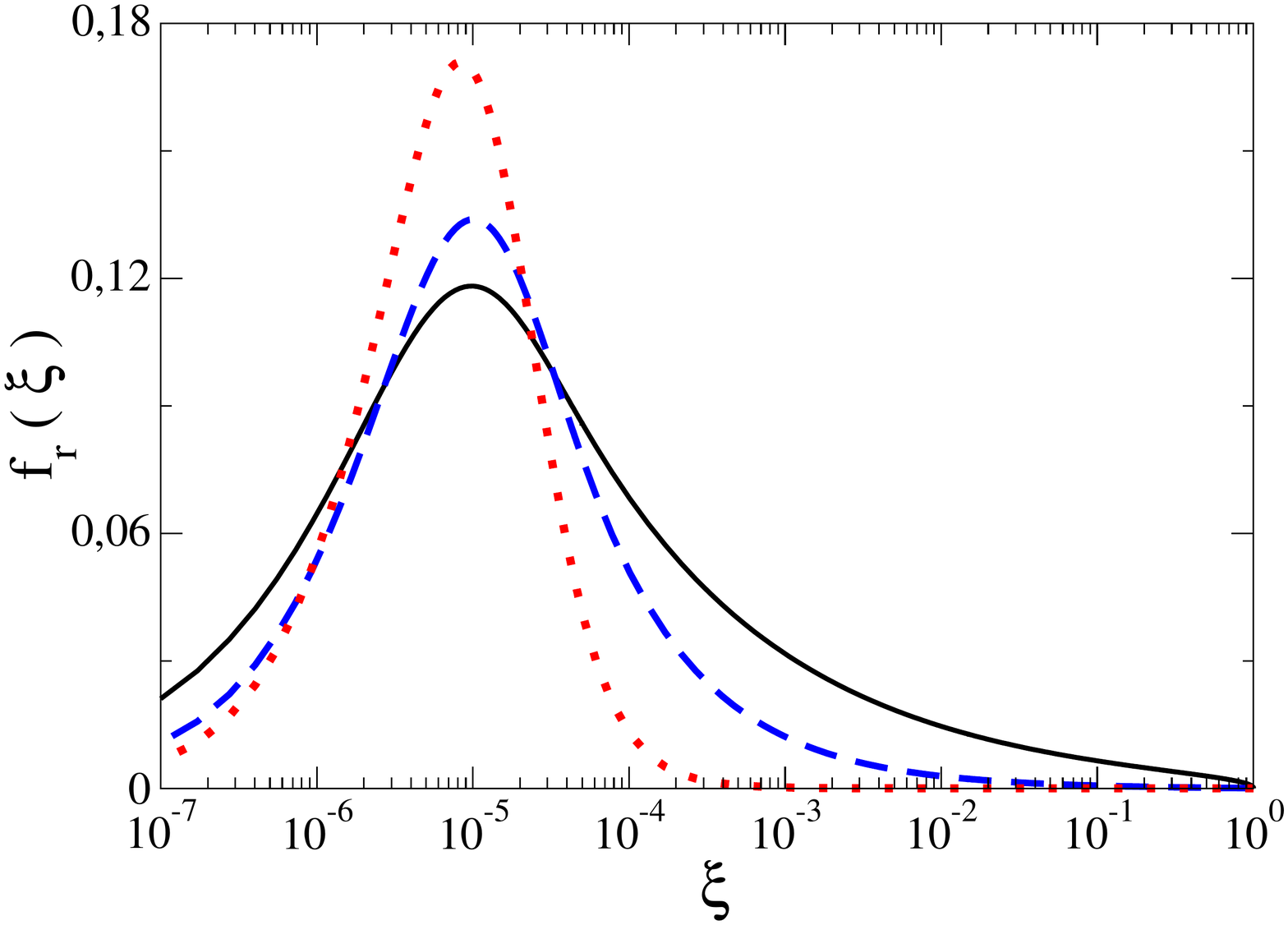,width=9.0cm}
\caption{(Color online). 
 $f_r(\xi)$ as a function of the momentum fraction $\xi$, for a set of the pair  $(|r|,\,\alpha)$ with an arbitrary normalization. Left panel: solutions  of Eq. \eqref{eq:BSfz2} in the Feynman gauge.    Solid line: $(|r_c|=0.455, \alpha_c = 1.187)$.   Dashed line: $(0.75, 0.8248)$. Dotted line: $(0.9999, 10^{-4})$. Right panel: the same as in the left panel, but for the Landau gauge. Solid line: $(|r_c|=0.522, \alpha_c = 1.276)$. Dashed line: $(0.75, 0.9811)$. Dotted line: $(0.9999,10^{-4})$. Note that $\xi=(1-z)/2$}\label{Fig2}
 \end{center}
\end{figure*}

To analyse the eigenfunctions $f_{r}(z)$ of Eq.~\eqref{eq:BSfz2}, it is convenient to use a new variable, $\xi = (1-z)/2$, that corresponds to the fermion longitudinal-momentum fraction. The function $f_{r}(\xi)$ for different values of $|r|$ and $\alpha$ is shown in Fig.~\ref{Fig2}. In the left panel, the calculation was performed in Feynman gauge, while, in the right panel, it was considered the Landau gauge.  For both gauges, $f_{r}(\xi)$ obtained  with the pair $(|r_c|,\alpha_c)$ is represented by  a solid line. From Fig.~\ref{Fig2}, one   can  notice that the general form of $f_{r}(\xi)$ is very similar in different gauges, and  it is vanishing at the end-points, as anticipated in the discussion of $f_r$ at $z=\pm 1$. In all cases, there is a peak at $\xi \sim 10^{-5}$, which means that the boson average  longitudinal-momentum fraction, $\langle 1-\xi \rangle $, is remarkably large.   Moreover, the results for the Feynman gauge, $\zeta=1$, are more close each other, while in the Landau gauge, $\zeta=0$, where only the transverse dofs of the vector-boson are present,  the differences are more pronounced. The sharpest peak can be obtained in  the Landau gauge in correspondence to $r=-1$, i.e. when the NWF fall-off is more fast.

 Let us summarize the salient features of $f_r(\xi)$: (i) the sharp peak for  $\xi$ close to 0  leads  to a   scalar diquark carrying almost all the {\em mock nucleon} longitudinal-momentum fraction; (ii) the Landau-gauge distributions have peaks  more  sharp than   the ones in  the Feynman gauge; (iii) in the two gauges, the distributions corresponding to $\alpha_c$ are quite similar, given  the small dependence of  the critical coupling on $\zeta$.
Interestingly, the property (i) entails that 
the massless fermion carries a spin opposite to the {\em mock nucleon} one,  since  for $\xi=0$ the fermion is moving toward negative $z$-axis and the massless fermion helicity is positive  (see also details in Ref.~\cite{AlvarengaNogueira:2019zcs}).
Then, the total spin of the composite state is obtained by adding one unit of orbital angular momentum, which is necessarily  of relativistic origin, as expected in the chiral limit. 

In the Feynman gauge, where the kernel contribution given in Eq.~\eqref{eq:G1Kernel} is acting, the height of the peak close to $\xi = 0$ is triggered by the maximum of the kernel contribution that appears for $z \sim z'$ (heuristically, one can  deduce this feature from the values of the ratios $(1\pm z)/(1\pm z')\le 1$ in combination with the corresponding theta-functions) and by the factor $ 4 \, |r| \, (1-z')^{-1}$,  which enhances the contribution for $z' \to 1$ (recall that $\xi=(1-z)/2$).
Analogously, the  feature (ii) in the Landau gauge, i.e.  the narrowest peak close to $\xi\to 0$ for $|r|> 0.5$, can be explained by analysing the kernel contribution given in  Eq.~\eqref{eq:G2Kernel}. The term  
$-|r| \, (1-2\, |r|) \, (1-z')^{-1}$ for $z>z'$ in Eq.~\eqref{eq:G2Kernel} 
is additive to the term $1+4|r|(1-z')^{-1}$  in Eq.~\eqref{eq:G1Kernel} when $|r|> 0.5$, enhancing the kernel for $z'\to 1$, and consequently resulting in a sharper peak of $f_r(\xi)$ in comparison to the Feynman gauge.
In this case, one has almost a Dirac delta for $|r| \to 1$ and one can assume a fermion {\em at rest}.

\subsection{Transverse degree of freedom}
We have  quantitatively investigated   the relation between what we have learned in  the chiral-limit analysis  of the BSE and the numerical solutions  of the ladder BSE for {\em massive} constituents and exchanged vector-boson, Eq.~\eqref{eqBSBF}, for large transverse momentum, (UV region). In this limit, all the masses can be disregarded with respect to $|\vec{k}_{\perp}|^2$ and the scale invariance becomes a very good symmetry. To proceed, let us define the components of the LF-projected BS amplitude~\cite{AlvarengaNogueira:2019zcs}
\be
    \psi_{i} (\gamma,z) = i \, M \int_{-\infty}^{+\infty} dk^{-} \phi_i(k,p) 
   \nonu
    = \int_{0}^{+\infty} d\gamma' \frac{g_i(\gamma', z)}{\left[\gamma' + \gamma + (1-z^2) \, \kappa^2 + z^2 \, \bar{m}^2 \right]^2} \, , 
\label{lfwf}
\ee
with $\gamma=|\vec{k}_{\perp}|^2 $.
From Eq.~\eqref{lfwf}, with the Ansatz given in Eq.~\eqref{eq:g2}, it is possible to conclude that the UV 
behaviour of the LF-projected BS amplitude is given by (recall that $r\in(-1,0)$)
\begin{equation}
    \psi_{2} ( \gamma,z) \underset{\gamma\to \infty}{\sim}  \gamma^{r-1}\, f_{r}(z) \, .
    \label{lfwfUV}
\end{equation}
The strategy to extract relevant information is to solve Eq.~\eqref{eqBSBF}, for a given set of parameters, and analyze if the solution  behaves as predicted by Eq.~\eqref{lfwfUV} for large momentum. In the Feynman gauge, the critical value of $\alpha_{c}$ is $1.187$ with the exponent $r_c = - 0.455$ (see Fig.~\ref{Fig1}). Therefore, in this  gauge, it is expected that the LF-projected BS amplitude corresponding to a coupling constant $\simeq \alpha_c$ decreases with $\gamma$, for any $z$, as follows  
\begin{equation}
    \psi_{2} (\gamma,z) \propto \gamma^{-1.455} \, .
    \label{lfwfUV2}
\end{equation}

\begin{figure}[thb]
\begin{center}

\epsfig{figure=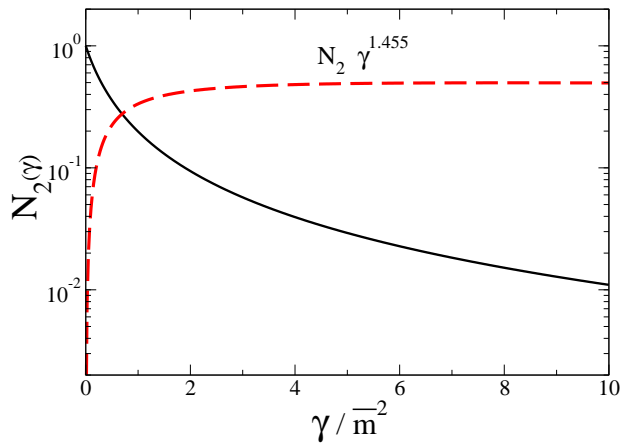,width=9.0cm}
\caption{(Color online). The normalized LF amplitude $N_2(\gamma)=\psi_2(\gamma,\xi_0 = 0.5)/\psi_2(0,\xi_0 = 0.5)$  as a function of the transverse momentum square $\gamma=|\vec{k}_{\perp}|^2$ in the Feynman gauge, for an equal-mass system, massless vector exchange and binding energy ratio $B/\bar{m} = 0.5$. Solid line: $N_2(\gamma)$. Dashed line: $ N_2(\gamma) \times \gamma^{1.455}$, with the power $r_c=-0.455$ corresponding to the critical value $\alpha_c=1.187$ (cf. Fig. \ref{Fig1}, left panel). }\label{Figextra}
 \end{center}
\end{figure}

In Fig.~\ref{Figextra}, it is presented the normalized LF-projected BS amplitude $$N_2(\gamma)={\psi_{2} (\gamma,\xi_0=0.5)\over \psi_{2} (0,\xi_0=0.5)}~,$$ obtained in Ref.~\cite{AlvarengaNogueira:2019zcs}, by numerically solving the Minkowski BSE, Eq.~\eqref{eqBSBF} for massive constituents, in the Feynman gauge for a massless vector exchange, and  equal-mass constituents, $m_f = m_s$. The  binding energy was chosen $B/\bar{m} = 0.5$, getting  $\alpha=1.189$, that is very close to its critical value. In Fig.~\ref{Figextra}, the dashed line is the product $N_2(\gamma)~\gamma^{1-r}$,  with $r=-0.455$, and  shows a clear asymptotic constant behavior. 

\begin{figure*}[hbt]
\begin{center}
\epsfig{figure=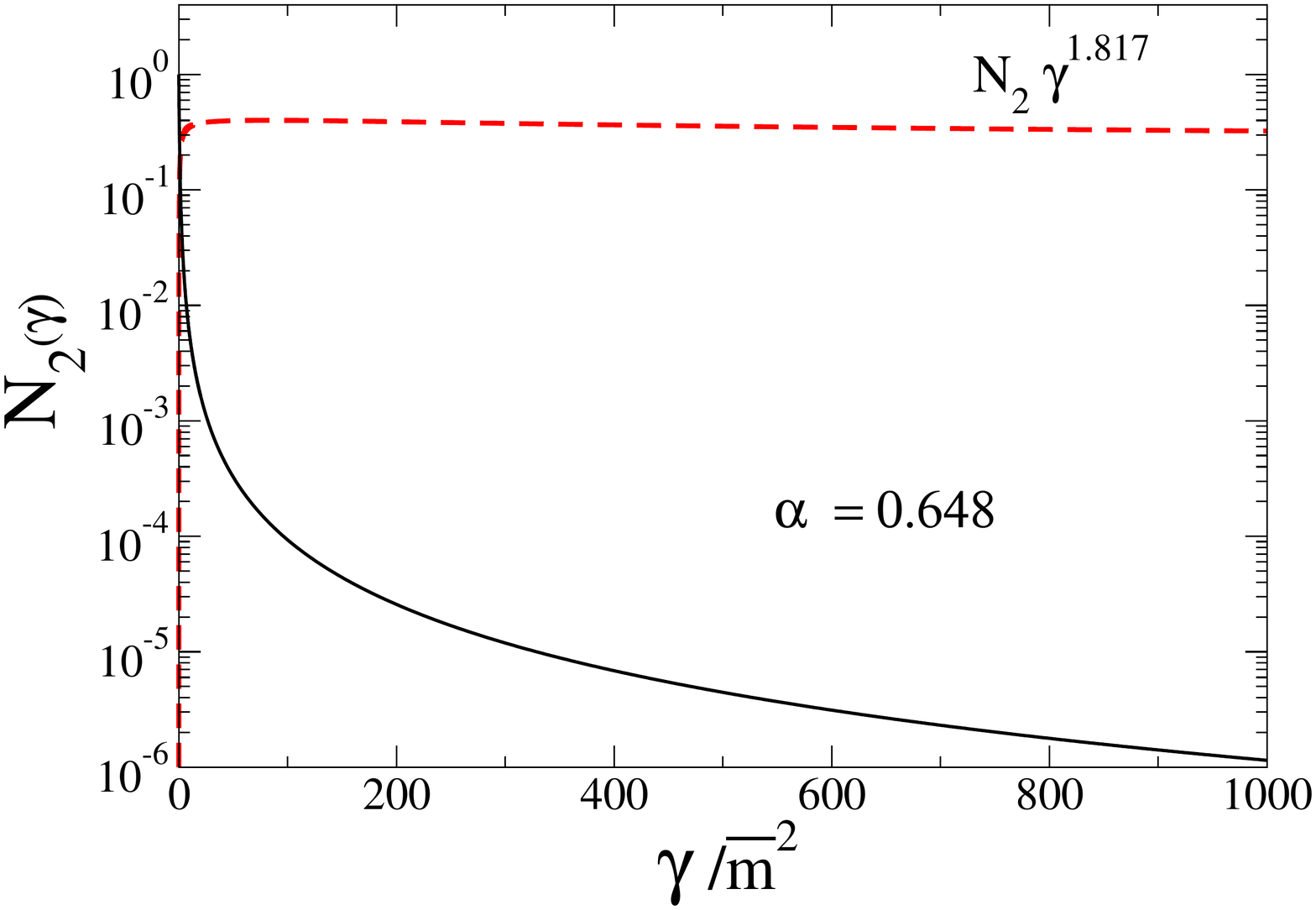,width=9.0cm}
\hspace{-0.5cm}\epsfig{figure=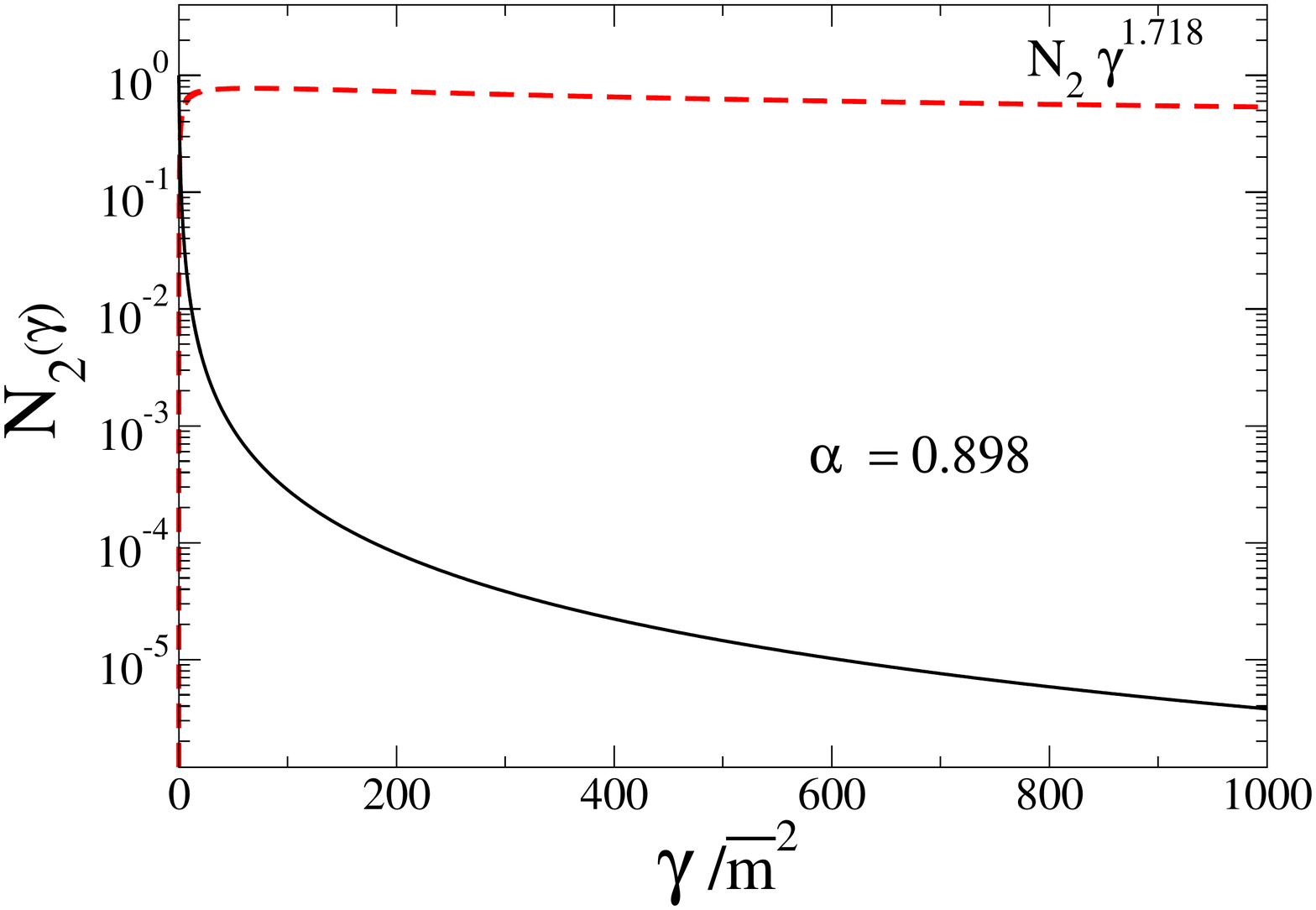,width=9.0cm}
\caption{(Color online). The same as in Fig. \ref{Figextra}, but  with $m_s/m_f = 2$ and binding energy ratio $B/\bar{m} = 0.1$ (i.e. $M/\bar m=1.9$). Left panel: $\mu/\bar{m}=0.15$ and $\alpha=0.648$. Solid line: $N_2(\gamma)$. Dashed line: $ N_2(\gamma) \times \gamma^{1.817}$.  Right panel: $\mu/\bar{m}=0.5$ and  $\alpha = 0.898$. Solid line: $N_2(\gamma)$. Dashed line: $ N_2(\gamma) \times \gamma^{1.718}$.}
\label{Fignucleon}
 \end{center}
\end{figure*}

In Fig. \ref{Fignucleon},  we present  two more comparisons in the Feynman gauge, that correspond to a {\em mock nucleon}, with $m_S/m_F=2$ and $M/\bar m=2 -B/\bar m=1.9$. The massive ladder-BSE is numerically solved in Minkowski space  with two different vector-boson masses: $\mu/\bar m=0.15$ and $\mu/\bar m=0.50$ (see Ref.~\cite{AlvarengaNogueira:2019zcs} for  details), obtaining $\alpha=0.648$ and $\alpha = 0.898$, respectively. For the first value of $\alpha$, one gets $r=-0.817$ from the physical branch of $\alpha(r,\zeta)$ in the left panel of Fig. \ref{Fig1}, while for the second value one has  $r=-0.718$. In the left panel, the calculation of $ N_2(\gamma)$, corresponding to $\alpha=0.648$ is shown together with the product $N_2(\gamma)~\gamma^{1-r}$ and in the right panel there is the comparison with the case $\alpha=0.898$. 

Notably, the constant behavior of the dashed lines  in Fig.  \ref{Figextra}  and \ref{Fignucleon} starts from values of $\gamma/\bar m^2$   order-of-magnitude different,    $\gamma/\bar m^2>6$ and  $\gamma/\bar m^2>600$, respectively. This is triggered by  the difference between the coupling constant and  its critical value, as one can realizes from   Fig. \ref{Figextra}, where   $\alpha=\alpha_c$ and  the scale invariance is established greatly earlier. Moreover, the striking constant behavior of the dashed lines {shown in Figs. \ref{Figextra}  and \ref{Fignucleon}} illustrates the predictive power of the chiral-limit analysis performed in this work, and suggests the possibility to extract quantitative signatures of the role of the one-particle exchange from the transverse-momentum distributions of hadrons.

\subsection{Longitudinal degree of freedom}
The analysis of the function $f_r(z)$ close to the end-point $z=-1$, i.e. a longitudinal-momentum fraction $\xi=1$, is interesting for the  insight one can gain on the valence parton distribution function, that should dominate the behavior close to $\xi=1$. 
Restricting to the Feynman gauge, for the sake of simplicity, one has the following approximation of  the integral equation in \eqref{eq:BSfz2} (cf. the kernel in Eq. \eqref{eq:G1Kernel}) for $z\to -1$
\bwt
\be 
f_r(z)\simeq {\alpha\over 2\pi} ~{1 \over 2 |r| (1-|r|)}
\Biggl\{\epsilon^{|r|}\int_{-1+\epsilon}^{1}\hspace{-0.3cm}dz'~\Biggl[{1 \over 1+z'}\Biggr]^{|r|} f_r(z')
 + 2^{|r|}
\int_{-1}^{-1+\epsilon}\hspace{-0.3cm}dz'~ \Biggl[{1\over 1-z'}\Biggr]^{|r|}\left(1 +{4|r|\over (1-z')}\right)\Biggr] ~f_r(z')\Biggr\}~,
\ee
\ewt
that leads to
\be 
 f_r(z) \to \epsilon^{|r|}\quad {\rm for} \quad z\to -1 +\epsilon~~. \ee
 Notice that the second integral  goes to zero more quickly that the first one, since $1> |r|$ and the interval of integration shrinks with the first power of   $\epsilon$.

From the above result one deduces that the valence parton distribution function goes to the end-point $\xi=1$ proportionally to $(1-\xi)^{2|r|}$, i.e. $(1-\xi)^{1.634}$
and $(1-\xi)^{1.436}$ for the two cases shown in Fig. \ref{Fignucleon}. It should be recalled that we are considering the chiral limit and a point-like scalar, and that in the limit $\alpha\to 0$ the exponent becomes 2. In general, our $1/2^+$ fermion-scalar bound system, in spite of the simple structure adopted, is able to  exhibit a large-$\xi$ behavior 
not too far from the exponent $3$ suggested by the counting rule (see, e.g.,  Refs.~\cite{Ball:2016spl,Accardi:2016qay,Bonvini:2019wxf,Hou:2019efy} for arguing an acceptable exponent as suggested by  the analysis of the proton experimental data).

\section{Summary}
\label{Sect:summary}
By using the Nakanishi integral representation of the Bethe-Salpeter amplitude, we have solved the homogeneous BSE in the chiral limit  for a $1/2^+$ bound system, composed by a fermion and a scalar, in Minkowski space. The interaction kernel is given by the one-particle exchange, with the actual dependence upon  the different covariant gauges. In the chiral limit,  one gets the decoupling of the two-channel integral system that determines the two Nakanishi weight functions, needed for the full reconstruction of the BS amplitude. Moreover,  by exploiting the scale invariance that establishes in the chiral limit one gets the factorization of the Nakanishi weight functions in an homogeneous function in $\gamma$ and  a function that depends upon the compact variable $z$.  Remarkably, the developed  formal analysis allows to determine the exact relation between the coupling constant, that drives the binding of the system, and the power that governs the power-like fall-off of the transverse-momentum amplitude, as well as the end-point behavior of the longitudinal momentum fraction amplitude.

Solutions have been presented for Feynman, Landau and covariant gauges in between, where the coupling constant has to be below a critical value in order to allow stable solutions, a problem already known for integral equations having symmetry under scale transformation. Above such critical coupling, the system breaks the continuous scale-invariant regime to a  discrete-scale one and the equations allow log-periodic solutions that demand the introduction of a boundary condition. This phenomena is associated with Efimov effect in non-relativistic few-body physics and to the Miransky scaling in the continuum framework of the quantum field-theory (see e.g.~\cite{Miransky:1984ef,Kaplan:2009kr}).
A striking qualitative feature in all analyzed covariant gauges, is that the scalar prefers to carry the nucleon momentum in the spin-antialigned fermion spin opposite to the system one) configuration. Therefore it is necessary to add the fermion-scalar orbital angular momentum to build the {\em mock-nucleon} spin, making clear the relativistic origin of such a configuration, already observed in Ref.~\cite{AlvarengaNogueira:2019zcs}.  

Finally, some relevant cases corresponding to both a massless vector-boson exchange and a massive one have been discussed, by comparing in the UV region  the solutions of the massive ladder-BSE and the chiral-limit prediction. The very good agreement suggests the interesting role of the hadron transverse-momentum distributions  in the search of quantitative  signatures of the one-particle exchange, that could be  investigated also analyzing  the end-point behavior of the valence parton distribution function (as it is well-known from the counting-rule predictions).
Plainly, a generalization of the present analysis to the Faddeev-Bethe-Salpeter equations is highly desirable,  also including the light-front approach (see, e.g., a first attempt in  Ref.~\cite{Ydrefors:2021mky,Ydrefors:2022bhq}), that is necessary  for addressing the phenomenology of the hadron momentum distributions.

\begin{acknowledgments}
This study was financed in part by Conselho Nacional de Desenvolvimento Cient\'ifico e Tecnol\'ogico (CNPq) under the grant 313030/2021-9, (WP), and 308486/2015-3 (TF), and  INCT-FNA project 464898/2014-5, and by Coordena\c c\~ao de Aperfei\c coamento de Pessoal de N\'ivel Superior (CAPES) under the grant 88881.309870/2018-01 (WP), and Funda\c c\~ao de Amparo \`a Pesquisa do Estado de S\~ao Paulo (FAPESP) Thematic grants 2017/05660-0 and 2019/07767-1.
\end{acknowledgments}
\appendix

%
\section{Kernel functions} 
\label{appendixa}
In this Appendix we present the expressions for the kernel of the integral equations (\ref{uvg1}) and (\ref{uvg2}). 
\begin{align}
\mathcal{P}_{1}^{(1)}({\bf V}) = - \int_{0}^{1} \, dv
\frac{\theta(z' - z)  \, (1+z)^{-1} \, \left(2 - v\right)}{\left[ \gamma \, (1-v)\, \left(\frac{1+z'}{1+z}\right) + \gamma'\right]^2}  \, . \nonumber\\
\label{P11UV1}
\end{align}
\bwt
\begin{align}
&\mathcal{P}^{(2)}_{1}({\bf V}) = - \frac12 \int_{0}^{1} dv  
\Biggl\{ 
 \frac{(1-v) \, (1+z)^{-1} \,  \theta(z'- z)}{\left[\gamma \, (1-v)\, \left(\frac{1+z'}{1+z}\right) + \gamma' \right]^3} ~ \left[\frac{(3- 5 v + 2 v^2) (1+ z') \,  \gamma}{(1+z)} + 3 (1-2 v) \,  \gamma'  \right] \, \Biggr\} \, .
\label{P11UV} 
\end{align}
\begin{align}
&\mathcal{P}_{2}^{(1)}({\bf V})
= - \int_{0}^{1} dv
\Biggl\{   \frac{ \theta(z' - z)}{ \gamma \left[] \gamma \, (1-v)\, \left(\frac{1+z'}{1+z} \right) + \gamma' \right]^2}  ~
\left[\gamma \, \frac{v (1-v) (1+z')}{2(1+z)} + \gamma' \, v \right] +  \frac{ \theta(z - z')}{\gamma \left[] \gamma \,  (1-v) \, \left(\frac{1-z'}{ 1-z} \right)  + \gamma'\right]^2}  \nonumber\\
& 
\times \left[  \gamma \, \frac{(1-v)(4 + v (1-z'))}{2 (1-z)} + \gamma' \, v \right]   \Biggr\} \, .
\label{P22UV1}
\end{align}
\begin{align}
&\mathcal{P}^{(2)}_{2}({\bf V}) = - \frac12 \int_{0}^{1} (1-v) \, dv  \Biggl\{ 
 \frac{\theta(z'- z)}{\gamma \left[\gamma \, (1-v)\, \left( \frac{1+z'}{1+z }\right) + \gamma' \right]^3} ~ \left[\gamma^2 \, \frac{(1-v)(1+z')(1+ v^2(1+z') - v(3+ 2 \, z'))}{(1+z)^2}  \right. \nonumber \\
& \left. 
+ \gamma \, \gamma' \, \frac{ (1-v (2+z')) }{(1+z)} + 3 \, \gamma'^2 \, v  \right] 
+ \frac{v \, \theta(z - z')}{\gamma \, \left[\gamma \, (1-v)\, \left( \frac{1-z'}{1-z} \right) + \gamma'\right]^3} \left[ - \gamma \, \gamma' \, \frac{(4-z')}{(1-z)} + 3 \, \gamma'^2  \right.  \nonumber\\
& \left.  
- \gamma^2 \, \frac{(1-v) (1-z') (1-v-z'(2-v))}{(1-z)^2} 
\right]  \, \Biggr\} \,  
\label{P22UV}
\end{align}
with ${\bf V}\equiv\{\gamma,z,\gamma',z'\}$.
\ewt
%
\section{Coupling constant $\alpha$ and power $r$}
\label{app_alpha}
This Appendix  illustrates the formal steps to obtain the relation between the  coupling constant $\alpha$ and the power $r$. The basic step is the integration on the variable $z$ of both sides in Eq. \eqref{eq:BSfz2}. 
In  particular, to perform the integration on the two kernel contribution in Eqs. \eqref{eq:G1Kernel} and \eqref{eq:G2Kernel}, respectively, the following result is useful
\bwt
\be
\int_{-1}^{+1}  dz \,  \left( {1\pm z \over 1\pm z'}\right)^{|r|} \, \theta(\pm z' \mp z) =  \left( 1\pm z' \right)  \,\int_{0}^{2/ (1\pm z')}  dy_\pm \,  y_\pm^{|r|} \, \theta(1 - y_\pm) 
= \,  \left( 1\pm z' \right)\int_{0}^{1}dy_\pm   \,   y_\pm^{|r|} = {1\pm z' \over 1+|r|} ~~,
\ee
with $
y_\pm = (1\pm z) / ( 1\pm z')$.
Then, one gets
\be
\int_{-1}^{+1}  G(z,z') \, dz  
= {1 \over 2 |r| (1-|r|)} \left[ {1+ z' \over 1+|r| } + \left( 1+ {4 |r|\over 1-z'} \right)  {1- z' \over 1+|r| } \right] 
= {1+ 2 |r| \over |r| (1-r^2)}~, 
\ee
and
\be
\int_{-1}^{+1}  H(z,z') \, dz 
=\frac{1}{2  |r|  (1-r^2)~(2-|r|) }  
~ \Bigg[  2  \frac{-|r|^3 + 7 \, |r|^2 - 11 \, |r| + 3}{ (3-|r|)} + r^2 \Bigg] 
 =\frac{   - 3 \, |r|^3 + 17 \, |r|^2 - 22 \, |r| + 6  }{2 |r|  (1-r^2)~(2-|r|) (3-|r|) }\, .  
\ee
Collecting the above result, from Eq.~\eqref{eq:BSfz2}
one gets
\be
C_0 = \int_{-1}^{+1}dz~ f_r(z) 
=\frac{\alpha}{2 \, \pi} \int_{-1}^{+1} dz'  \int_{-1}^{+1} dz ~\Bigl[ {G}(z,z') 
+ (1-\zeta)  {H}(z,z')\Bigr]  \,  f_{r}(z') \nonu
= \frac{\alpha}{2 \, \pi} \Biggl[ {1+ 2 |r| \over |r| (1-r^2)} 
 + (1-\zeta)  \frac{ - 3 \, |r|^3 + 17 \, |r|^2 - 22 \, |r| + 6  }{2\, \, |r| \, (1-r^2)~(2-|r|) (3-|r|) }   \Biggr] 
 \, \int_{-1}^{+1} dz'    f_{r}(z') \, ,
\ee
and therefore
\be
\alpha =   2 \pi \,  \Biggl[ {1+ 2 |r| \over |r| (1-r^2)} 
 + (1-\zeta)  \frac{ - 3 \, |r|^3 + 17 \, r^2 - 22 \, |r| + 6  }{2\, \, |r| \, (1-r^2)~(2-|r|) (3-|r|) }   \Biggr]^{-1}.
\label{eq7}
\ee
\ewt

\end{document}